\documentclass{emulateapj}
\usepackage{amsmath}
\usepackage{epsfig}
\usepackage{amssymb}
\usepackage{graphicx}
\usepackage{color}
\usepackage{natbib}
\usepackage{soul}
\usepackage{cancel}

\usepackage{apjfonts} 
\usepackage{amsmath,amstext}
\usepackage[breaklinks,colorlinks,citecolor=blue,linkcolor=magenta]{hyperref} 
\usepackage[all]{hypcap} 
\usepackage{verbatim}
\usepackage[dvipsnames]{xcolor}

\newcommand{\HI}{H{\sc \ i}}

\begin{document}

\title{An estimate of the impact of reionization on supermassive black hole growth}

\author{Phoebe R. Upton Sanderbeck$^1$, Jarrett L. Johnson$^1$ \& Madeline A. Marshall$^1$}
 \email{phoebeu@lanl.gov}
 \affiliation{1. Los Alamos National Laboratory, Los Alamos, NM, USA}

\date{\today}

\begin{abstract}
The supermassive black holes (SMBHs) that power active galactic nuclei found at $z\geq 6$ were formed during the epoch of reionization. Because reionization is an inhomogeneous process, the physical properties of SMBH host galaxy environments will vary spatially during reionization. We construct a semi-analytic model to estimate the impact of reionization on SMBH growth. Using a series of merger trees, reionization models, and black hole growth models, we find that early reionization can reduce an SMBH's mass by up to [50, 70, 90] \% within dark matter halos of mass [$10^{12}$, $10^{11}$, $10^{10}$] M$_{\odot}$ by $z$ = 6. Our findings also suggest that the redshift range in which black hole growth is impacted by reionization strongly depends on whether the Eddington accretion rate can be exceeded. If so, we find that black hole masses are significantly suppressed principally during the early phases of reionization ($z$ $\ga$ 10), while they are more readily suppressed across the full redshift range if super-Eddington growth is not allowed. We find that the global average impact of reionization may be to reduce the masses of black holes residing in $\la$ 10$^{11}$ M$_{\odot}$ halos by a factor of $\ga$ 2. The census of supermassive black holes being uncovered by the {\it James Webb Space Telescope} may offer a means to test the basic prediction that more massive black holes reside in cosmological volumes that are reionized at later times.

\end{abstract}

\maketitle

\section{Introduction}

The reionization of hydrogen, the last major phase change in the Universe, occurred when the first galaxies produced sufficient ultraviolet photons to break apart the bulk of the hydrogen atoms in the intergalactic medium into their constituent protons and electrons. Reionization induced a significant temperature increase in intergalactic gas that accompanied the change in ionization state. Because reionization is an inhomogeneous process, the ionization and temperature fields varied spatially throughout the cosmos during this epoch. 

Reionization is thought to end around $z\sim 5-6$ \citep{fan06,kulkarni19,keating20,nasir20}, and the number of quasars known to exist at these redshifts and higher has been steadily growing. With more than 250 quasars discovered at $z>6$ \citep{Mortlock2011, Banados2018, inayoshi19, bosman20}, several hosting $>10^9$ M$_{\odot}$ black holes \citep{yang20}, much of the growth of these active galactic nuclei (AGN) must have occurred during the epoch of reionization. Supermassive black holes (SMBHs) driving these quasars are assembled within a Gyr of the Big Bang -- a fact that is in contention with the paradigm of Eddington-limited accretion onto massive Pop III seed black holes \citep{VR2006, Alvarez2009, Tanaka2009, Smith2018}. Mechanisms such as super-Eddington growth, growth through frequent mergers, and more massive black hole seeds can help relieve this tension. However, reionization could impede rapid growth at later times by limiting the gas supply available to a growing SMBH. 

The reionization of hydrogen suppresses the growth of small galaxies through the increase in Jeans mass of the intergalactic gas that results from its photoheating \citep{couchman86,shapiro94,quinn96,thoul96,bullock00,gnedin00,wyithe06,hoeft06,okamoto08,sobacchi13,noh14}. Additionally, ionizing radiation affects the rate at which gas can cool \citep{efstathiou92,dijkstra04,hambrick09,hambrick11}, and thus accrete. Accretion of intergalactic gas can be suppressed onto dark matter halos as large as $\sim$ $10^{11}$M$_{\odot}$ at $z=0$ due to the photoionization of the gas from reionization \citep{noh14}. Though the bright AGN found at $z>6$ are hosted in halos that have virial temperatures that far exceed the temperature to which reionization heats the intergalactic medium (IGM) ($T_{\rm reion}\approx 17,000-30,000$, e.g. \citealt{daloisio19}), the early stages of the SMBH's growth may have occurred when the host halo was below this threshold. Finally, the initial size of seed black holes can be limited in host halos that experience reionization early in their growth \citep{johnson2014, chon2017, agarwal2019}.

Recent evidence has suggested that SMBHs may assemble more rapidly at early times (e.g., \citealt{Lai2024}). Observations by the {\it James Webb Space Telescope} (JWST) have revealed numerous SMBHs that have been assembled rapidly as early as $z$ $\sim$ 10, particularly with respect to the stellar mass in their host halos \citep{Natarajan2023, Pacucci2023, Scoggins2023, Stone2023}. And while SMBHs as large as $1.6 \pm 0.4 \times 10^{9}$ M$_{\odot}$ at $z=7.64$ \citep{wang21} and $1.2\times 10^{10}$ M$_{\odot}$ at $z=6.3$ \citep{wu15} have been discovered in high redshift quasars, at low redshift the largest SMBHs found are no more than an order of magnitude more massive than what has been found at $z>6$. While their growth could be limited by overwhelming feedback at these large masses \citep{pacucci17} or accretion disk instability \citep{Inayoshi2016}, a dearth of neutral intergalactic gas may also result in slower growth at late times. The early emergence of these SMBHs raises the question: Are the conditions in the reionization-era Universe allowing early SMBHs to assemble so quickly?

In addition to the most massive SMBHs inferred to power high-z quasars during the epoch of reionization, JWST has now uncovered numerous lower-mass BHs in AGN at these same high redshifts \citep{Greene2023,Larson2023,Harikane2023,bogdan23,Kocevski2023,maiolino23,Matthee2023,Juodzbalis2023,Ubler2023}. These JWST discoveries have not yet remedied the tension between current models of black hole growth and the emergence of SMBHs at these redshifts; for instance, \citet{Jeon2023} find that a Bondi-Hoyle prescription does not yield a match to data in models of low-luminosity AGN detected by JWST. These JWST AGN have also shed light on an unexpected ubiquity of quasars at high redshift \citep{Greene2023}. \citet{Juodzbalis2023} find an AGN fraction of at least 5 percent between $z=6.5$ and $z=12$.

Also puzzling is the presence of neutral gas around certain known high-redshift quasars. If these quasars reside in the rare, steep cosmic density peaks in which the first galaxies cluster, it should be expected that these regions are reionized earliest. However, there is as of yet no strong observational consensus to support this  \citep{Kim2009, Mazz2017, fan19}, leaving open the possibility that the black holes powering these bright high-z quasars grow in neutral regions of the Universe during the early stages of reionization.

Here we consider the impact of the local reionization history on the growth of high-z quasars and AGN, accounting for a reduced gas supply that is expected to persist in regions that experience reionization at relatively early times.
While high-redshift quasars are often used as observational tools with which to study reionization \citep{Fan2022}, here we are thus using models of reionization to understand the properties of high-redshift quasars and the growth of the SMBHs that power them.

  The impact of other feedback processes on SMBH growth on the galactic scale have previously been investigated; for example, numerical simulations suggest that supernova feedback impedes SMBH growth in low-mass halos \citep{dubois15,prieto17,habouzit17,trebitsch18}. Similarly, \citet{boothandschaye} find that supernova feedback weakens AGN activity. \citet{daniel17}, however, find that black hole feedback does not change the black hole mass - stellar mass relation in their simulations. While these feedback effects pertain to the galactic scale, here we focus for the first time on how the extragalactic feedback from reionization affects the overall gas available for black hole growth.

 In Section~\ref{sec:model}, we present our semi-analytic approach to modeling the effect of reionization on BH growth.  We present our results in Section~\ref{sec:results} and discuss these results as well as observational prospects and potential measurements in Section~\ref{sec:obs}. Finally, we provide a summary of our findings in Section~\ref{sec:summ}.

\section{Semi-analytic model}
\label{sec:model}
\begin{figure}
\resizebox{8.7cm}{!}{\includegraphics{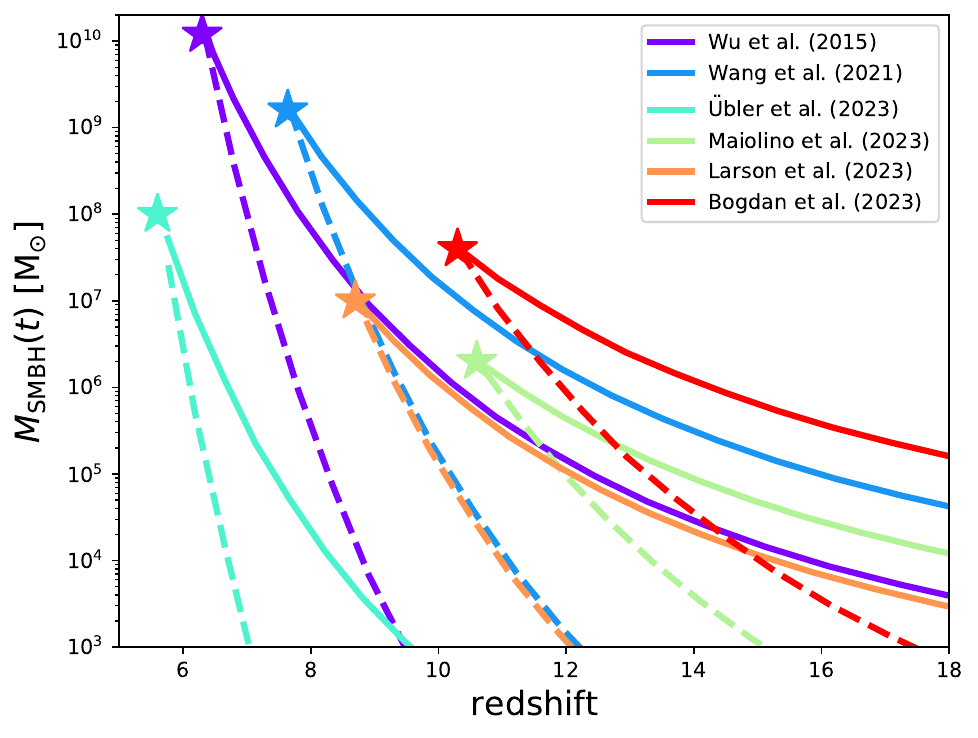}}\\
\caption{Growth histories during the epoch of reionization ($z$ $\ga$ 5.5) of known SMBHs, assuming accretion at the Eddington rate ({\it solid lines}) and twice the Eddington rate ({\it dashed lines}),  with a radiative efficiency of $\epsilon$ = 0.1. }
\label{fig:edd}
\end{figure}

Before we describe our model, a simple demonstration of the concurrence of early SMBH growth with reionization is shown in Figure~\ref{fig:edd}. Here we show hypothetical growth histories of several black holes inspired by recent SMBH discoveries by JWST at high redshift \citep{bogdan23,Larson2023,maiolino23,Ubler2023} as well as two previously discovered SMBHs that were novel discoveries in regards to their masses and redshift \citep{wu15,wang21}. Figure~\ref{fig:edd} shows the growth histories of the aforementioned SMBHs, assuming accretion at the Eddington rate (solid curves) and at double the Eddington rate (dashed curves). In all cases we adopt a standard radiative efficiency of $\epsilon = 0.1$. The span of the x-axis represents a rough approximate window of the timing of reionization. Figure~\ref{fig:edd} demonstrates that the majority of SMBH growth for these objects must have occurred during or prior to reionization.

In the last several decades, there has been a wealth of literature investigating the impact of reionization on the gas content of galaxies (e.g. \citealt{couchman86,efstathiou92,shapiro94,quinn96,bullock00,gnedin00,dijkstra04,wyithe06,hoeft06,okamoto08,sobacchi13,noh14}). These studies find that, by and large, low mass galaxies are able to retain only a fraction of the gas that they would otherwise be able to retain in the absence of reionization. Though many of the quasars so far observed at high redshift reside in host galaxies larger than those most affected by reionization, hierarchical growth dictates that these larger host halos were assembled by the coalescence of smaller halos, many of which should have had a reduced or negligible gas fraction due to reionization.

In the remainder of this section, we describe the semi-analytic framework that we use to model changes in SMBH growth due to reionization. We first consider how reionization alters gas content in small halos, and construct merger trees to assess how the population of merging galaxies is impacted as a result. We then estimate the impact this has on black hole growth for a variety of reionization redshifts, as well as for an extended, inhomogenous reionization model. Finally, we explain the model in full -- we estimate the change in gas accretion onto a SMBH in a halo of a given mass due to the suppression of gas-rich mergers.

\subsection{The suppression of accretion onto halos due to reionization}

To understand how reionization affects the gas content in halos, one must consider the interweaving of several physical processes that include gravity, heating, cooling, and self-shielding. However, the simplest way to approximate the effect of reionization on gas content in a galaxy is to make a basic Jeans mass argument. The Jeans mass, which can be derived from the virial theorem, is dependent on the temperature of the gas and ionization state such that
\begin{equation}
M_{J} = 4\times 10^9 \left( \frac{T}{10^4 {\rm K}} \right)^{3/2} \left(\frac{n_{\rm H}}{10^{-3} {\rm cm^{-3}}}\right)^{-1/2} {\rm M}_{\odot}
\end{equation}
 where $T$ is the temperature, and $n_{\rm H}$ is the number density of hydrogen atoms. However, simulation results show that the Jeans mass is not a good prescription for the halo mass above which gas can accrete, as calculations that holistically consider redshift, the thermal history of intergalactic gas, the growth history of halos, and the ionizing background are more successful at reproducing the trends found in simulations \citep{noh14}. Most of these models find that reionization most successfully suppresses gas in halos with masses less than $\sim 10^8$ M$_{\odot}$ up to high redshifts, with halos of increasing mass experiencing suppression at lower redshifts.

Despite advances in understanding how reionization suppresses gas content in small galaxies, many of the models are constructed with a specific halo collapse redshift and reionization model. Due to the varying reionization redshifts and uncertainties associated with the redshift at which a halo collapses in our model, we adopt a less sophisticated approach than these previous works to capture the effect of reionization on gas content in small halos. For simplicity, we adopt a redshift-dependent threshold mass below which halos will be unable to retain gas following reionization. 

We use two physically-motivated criteria to determine this redshift-dependent mass threshold:
\begin{enumerate}
    \item That for which the corresponding virial temperature of the halo is equal to $\ga$ 1.0$\times10^4$ K or $\ga$ 2.5$\times10^4$ K. These temperatures aim to bracket the temperatures at which reionization may heat the local intergalactic gas as well as the temperature that may be retained by the gas following reionization. 
    \item One quarter of the Jeans mass ($1/4$M$_{J}$) evaluated at 1.0$\times$10$^4$ K and the turnaround density.
\end{enumerate}

The virial temperature in this case is \citep{barkana01},
\begin{equation}
T_{\rm vir} = 2 \times 10^4 \left( \frac{\mu}{0.6} \right)\left( \frac{M}{10^8 h^{-1} M_{\odot}}\right)^{2/3} \left[ \frac{\Omega_{m}}{\Omega_{m}^{z}}\frac{\Delta_{c}}{18\pi^2}\right]^{1/3}\left(\frac{1+z}{10}\right) {\rm K},
\end{equation}
where $\mu$ is the mean molecular weight and $\Delta_{c}$ is the overdensity at the collapse redshift (estimated to be 18$\pi^2$ in an Einstein-de Sitter Universe). This simple prescription for the threshold halo mass assumes that gas cannot collapse into a halo if it is photoheated to temperatures above the virial temperature of the halo. Though this criterion lacks the sophistication of many of the numerical studies, it does provide a more conservative estimate than many of the more physically-motivated models. The bracketing values of virial temperature we use are consistent with those found in \citet{Weer2023}, who model dwarf galaxies and choose virial velocity cutoffs of 19 and 25 km s$^{-1}$ .

For the remaining model, we adopt a threshold mass defined by one fourth of the maximum Jeans mass. This model is the most physically motivated, as \citet{noh14} found this to be a reasonably accurate condition for gas accretion in their cosmological hydrodynamics simulations. Here we adopt the turnaround density and a temperature of 1.0$\times10^4$ K.

\begin{figure}
\resizebox{9.0cm}{!}{\includegraphics{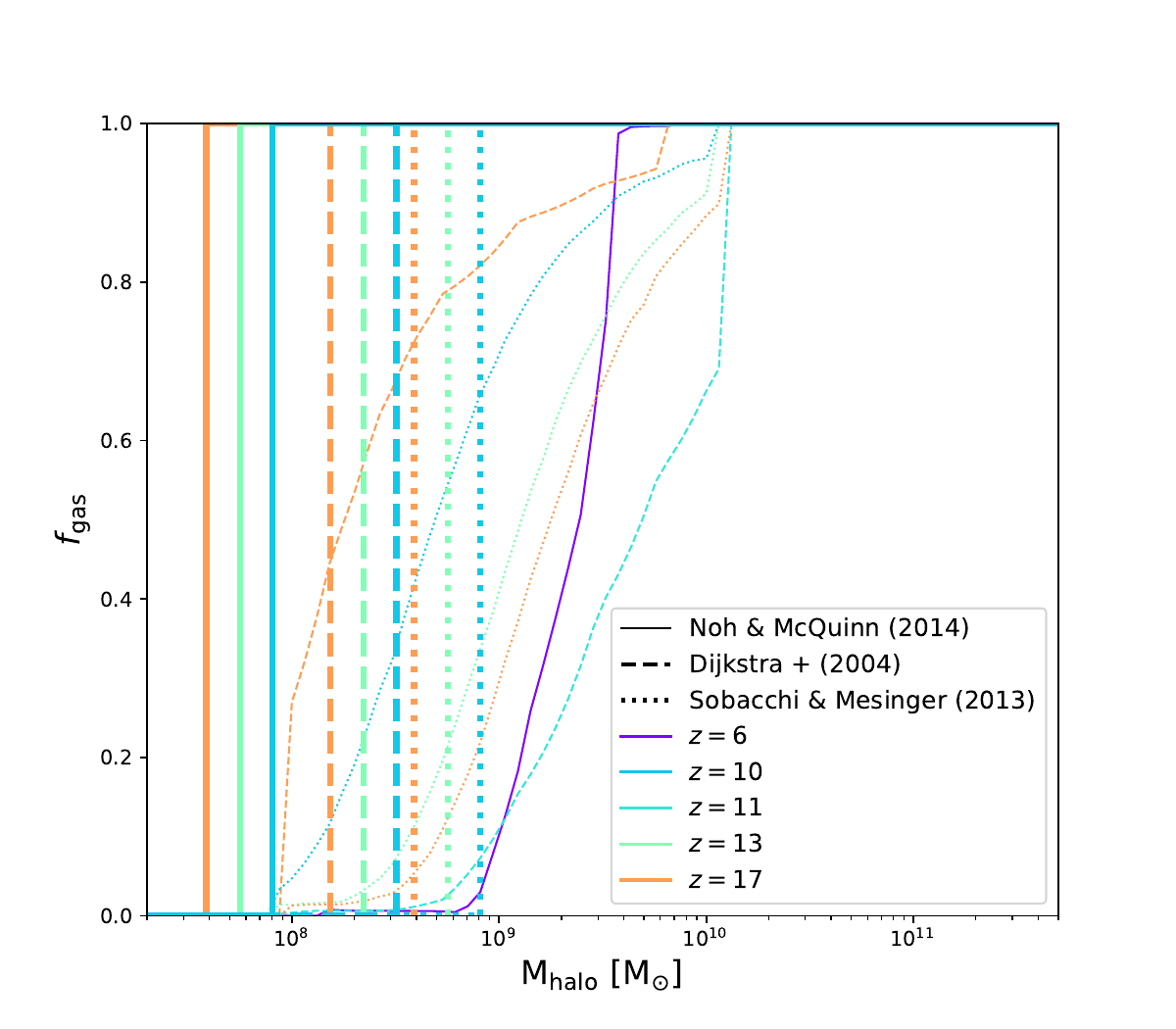}}\\
\caption{ Gas fractions as a function of halo mass. The thick vertical lines are the models that we adopt, where the dotted vertical lines show a threshold mass defined by a quarter of the Jeans mass at a gas temperature of $1.5 \times 10^4$ K, the dashed vertical lines show the threshold mass determined by a viral temperature of $2.5 \times 10^4$ K, and the solid vertical lines show the threshold mass determined by a viral temperature of $10^4$ K. The corresponding redshifts are selected to match the historical models shown from \citet{dijkstra04} and \citet{sobacchi13}. We additionally include a $z=6$ model from \citet{noh14}.
\label{fig:gfrac}}
\end{figure}

Between our three conditions for determining the threshold mass -- a virial temperature of 1.0$\times10^4$ K,  a virial temperature of 2.5$\times10^4$ K, and $1/4$M$_{J}$ -- we aim to bracket the plausible range of masses for which reionization could suppress gas accretion at the redshifts we consider. Figure~\ref{fig:gfrac} shows our three choices of the threshold halo mass at $z=17$, $13$, and $11$, represented by the vertical lines. The solid vertical lines represent the most conservative of our thresholds (a virial temperature of 1.0$\times10^4$ K), the dashed represent the fiducial model (a virial temperature of 2.5$\times10^4$ K), and the dotted vertical lines represent the $1/4$M$_{J}$ threshold. Figure~\ref{fig:gfrac} includes three models from historical literature, specifically from \citet{dijkstra04,sobacchi13}; and \citet{noh14}, for comparative purposes. While our threshold masses present a more simplistic approach to modeling the impact of reionization on gas content in halos, they mostly provide a more conservative estimate of the effect than these models.

\subsection{Merger tree model}
The statistical history of a halo's growth through time can be described analytically through a merger tree model based on an extended Press-Schechter formalism (e.g.,\citealt{lacey93,somerville99}). Modern analytic merger tree models are found to successfully reproduce halo assembly in N-body simulations, capturing the growth of the main progenitor halo and the statistics of the merging halos.

We use the open-source code \textsc{Galacticus} \citep{galacticus} for our merger tree calculations. \textsc{Galacticus} is a flexible, semi-analytical galaxy formation model that grows galaxies through a merging hierarchy of dark matter halos in a dark matter-dominated universe. \textsc{Galacticus} has been recently used to model the formation of Milky Way satellite galaxies \citep{Weer2023} as well as SMBH populations \citep{Liempi2023}.

\begin{figure}
\resizebox{9.0cm}{!}{\includegraphics{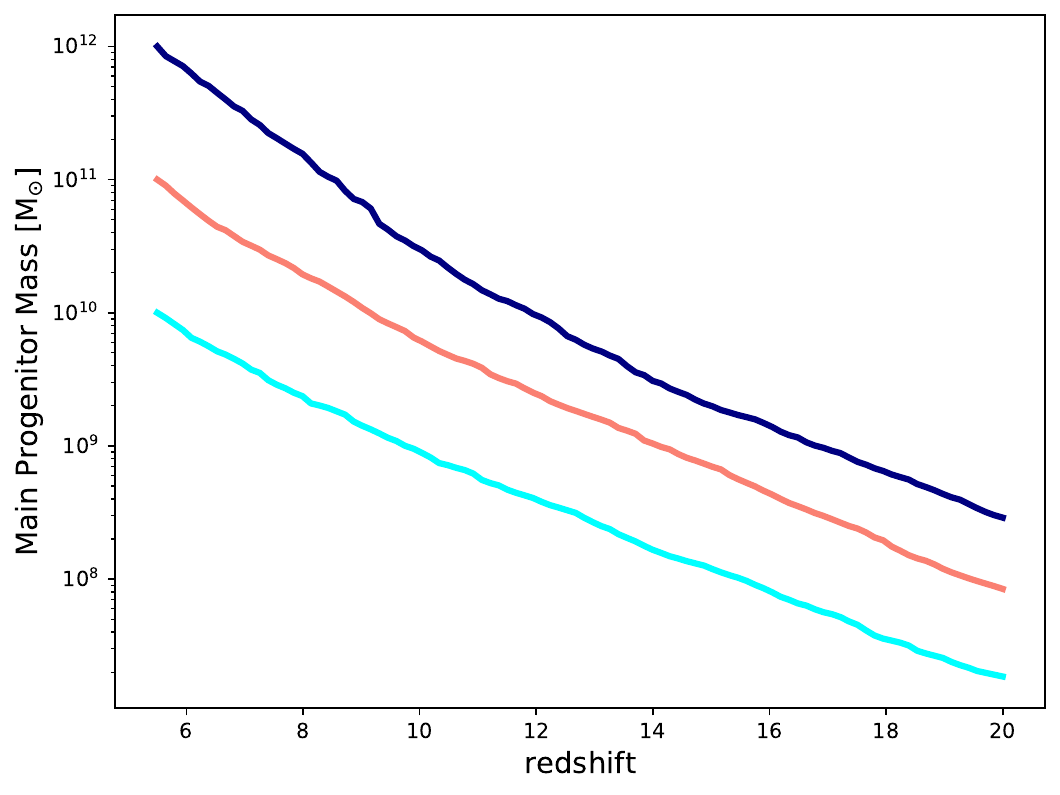}}\\
\caption{The growth of the main progenitor halos in our merger trees. 
\label{fig:MP}}
\end{figure}

We calculate a series of merger trees from $z\approx20$ down to $z=5.5$ for three different final halo masses: $10^{12}$, $10^{11}$, and $10^{10}$ M$_{\odot}$. For each final halo mass, we run a series of twenty trees with different random seeds. To capture a sufficient halo mass resolution, we set our minimum halo mass to $10^6$ M$_{\odot}$. Figure~\ref{fig:MP} shows the average growth of the main progenitor halos in our series of merger tree calculations. Each of these growth histories is the mean of the twenty merger tree realizations.

\begin{figure*}
\begin{center}
\resizebox{18.5cm}{!}{\includegraphics{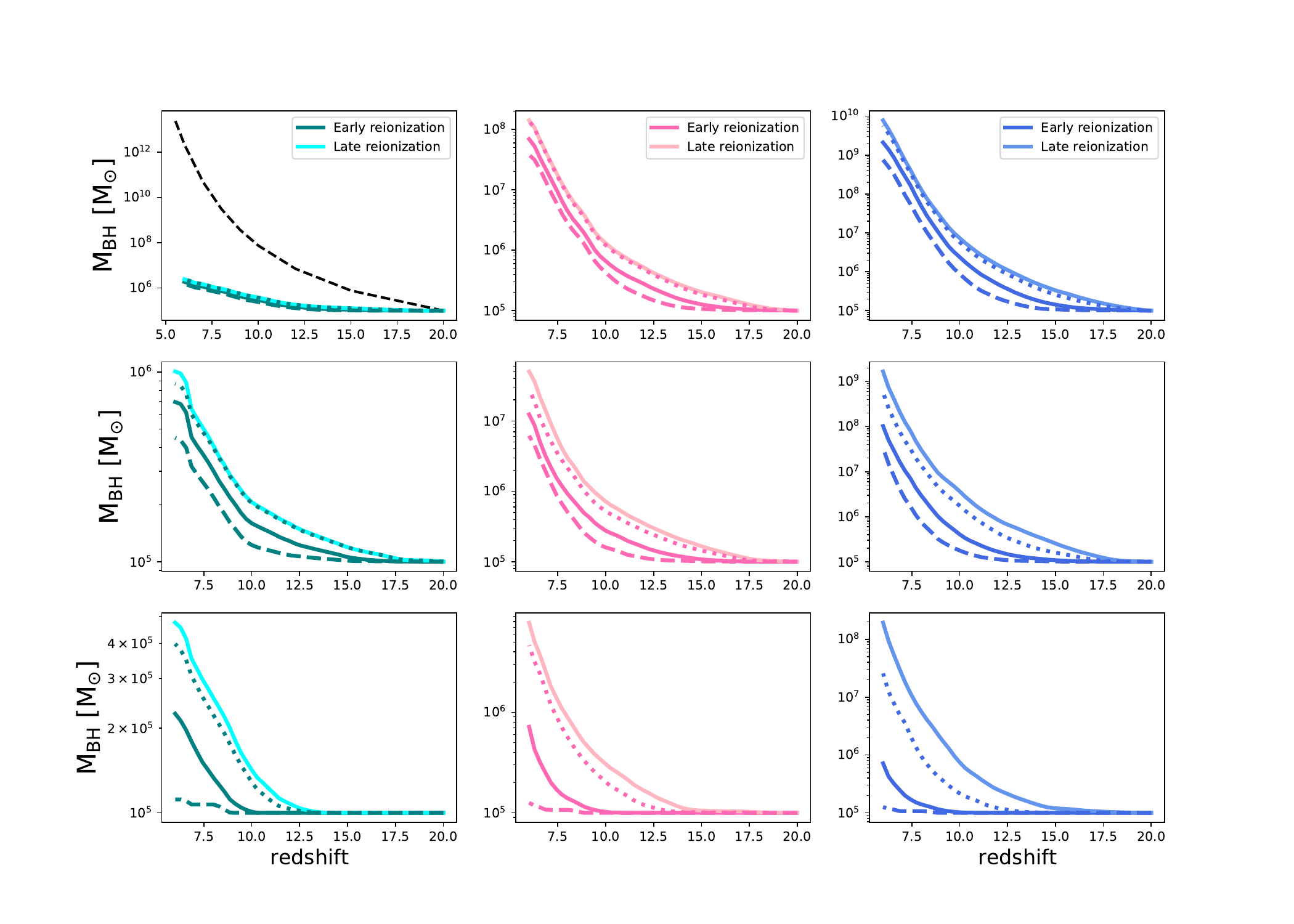}}\\
\end{center}
\caption{The growth histories of SMBHs in our Eddington-limited model. The lighter curves show the growth preceding a late reionization and the darker curves show growth following an early reionization. The dotted curves show an early reionization model with our weakest gas suppression condition, a virial temperature of 1.0$\times 10^4$ K. The solid darker curves show the same, but with a virial temperature of 2.5$\times 10^4$ K. The lower dashed curves show a calculation with our threshold halo mass that can retain gas following reionization as defined by one fourth of the Jeans mass. The upper panels show a final halo mass of $10^{12}$ M$_{\odot}$ at $z=5.5$, the middle panels show a final halo mass of $10^{11}$ M$_{\odot}$ at $z=5.5$ and the lower panels show a show a final halo mass of $10^{10}$ M$_{\odot}$ at $z=5.5$. The left panels show results for the case in which a "major merger" that triggers BH growth is defined as the merging halo possessing one third of the main progenitor's gas mass, the middle panels show this ratio as one tenth, and the right panels show the case of a relaxed threshold of one twentieth. The black dashed line in the top-left panel shows the growth history under the assumption of constant accretion at the Eddington rate, for reference. 
\label{fig:result1}}
\end{figure*}

\begin{figure*}
\begin{center}
\resizebox{18.5cm}{!}{\includegraphics{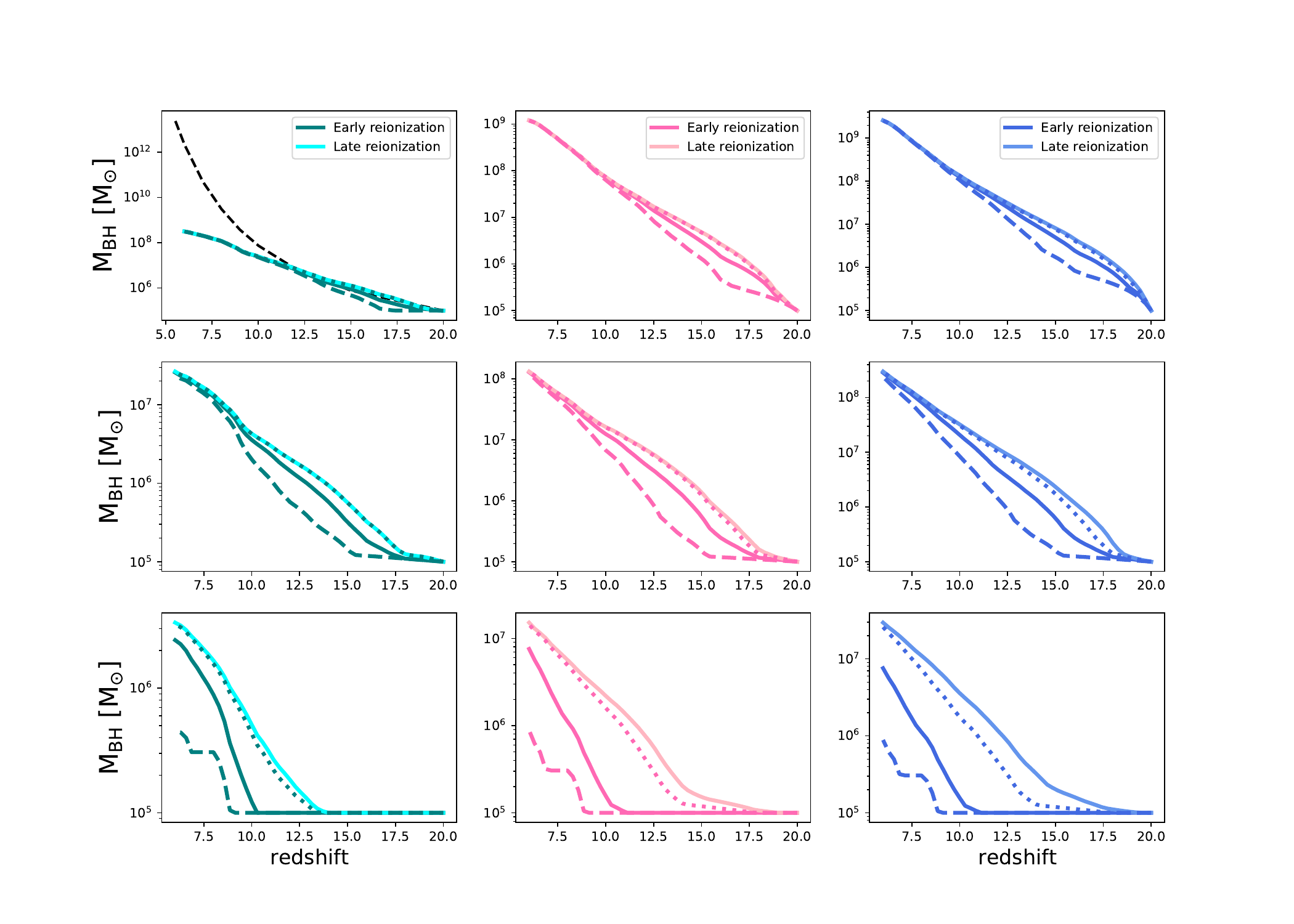}}\\
\end{center}
\caption{Just as Figure~\ref{fig:result1} but for our non-Eddington-limited model. 
\label{fig:result2}}
\end{figure*}

\begin{figure*}
\begin{center}
\resizebox{20.0cm}{!}{\includegraphics{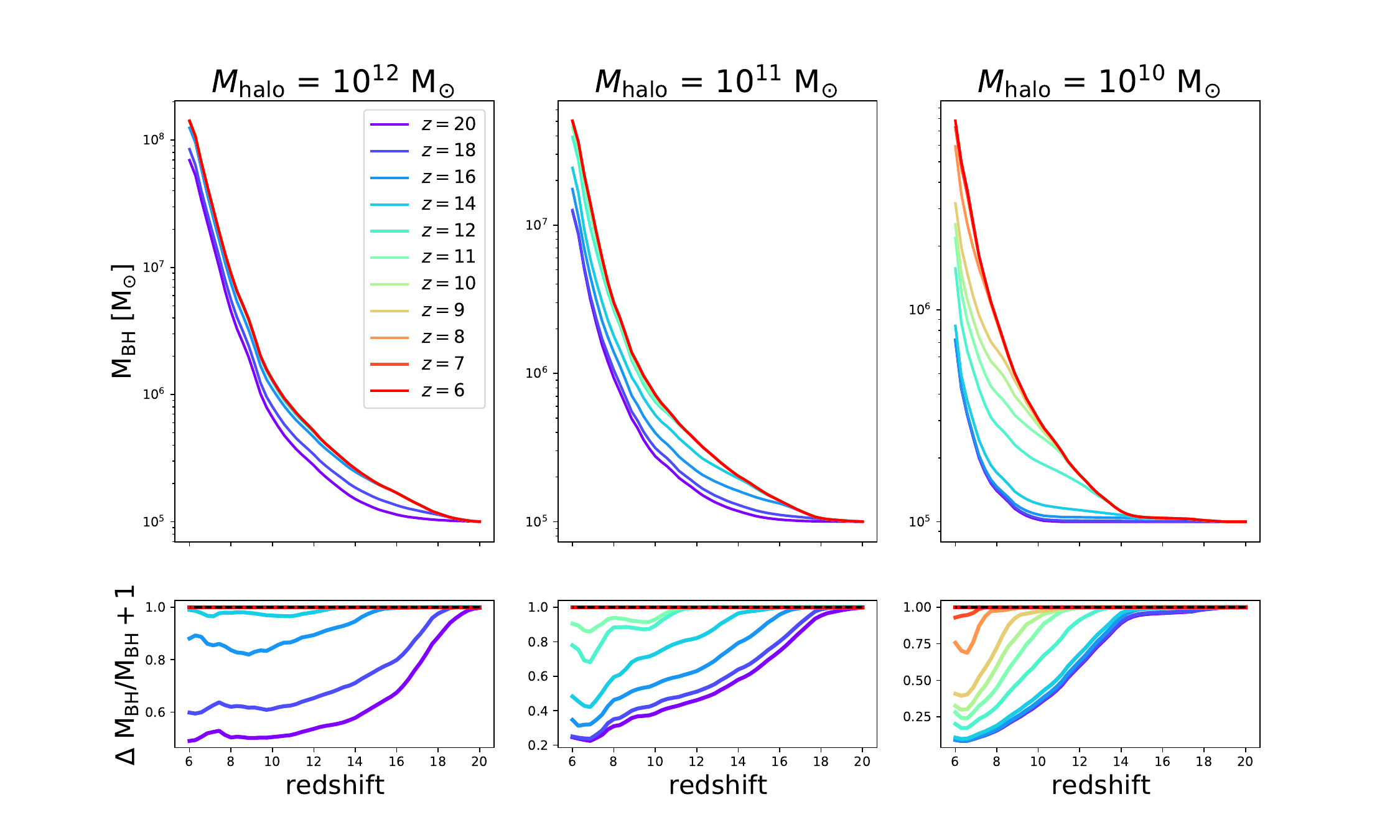}}\\
\end{center}
\caption{The growth histories of SMBHs in our Eddington-limited model experiencing different local reionization redshifts. The lower panels show the fractional change in SMBH mass between the late reionization ($z$ = 6) models and the earlier reionization models in the upper panels. From left to right, the panels show results for a final halo mass of $10^{12}$ M$_{\odot}$, 10$^{11}$ M$_{\odot}$ and 10$^{10}$ M$_{\odot}$ at $z=5.5$. In these calculations, we assume that the gas mass of a "major merger" that triggers BH growth is defined as ten percent of the main progenitor's gas mass. }
\label{fig:result3}
\end{figure*}

\begin{figure*}
\begin{center}
\resizebox{20.0cm}{!}{\includegraphics{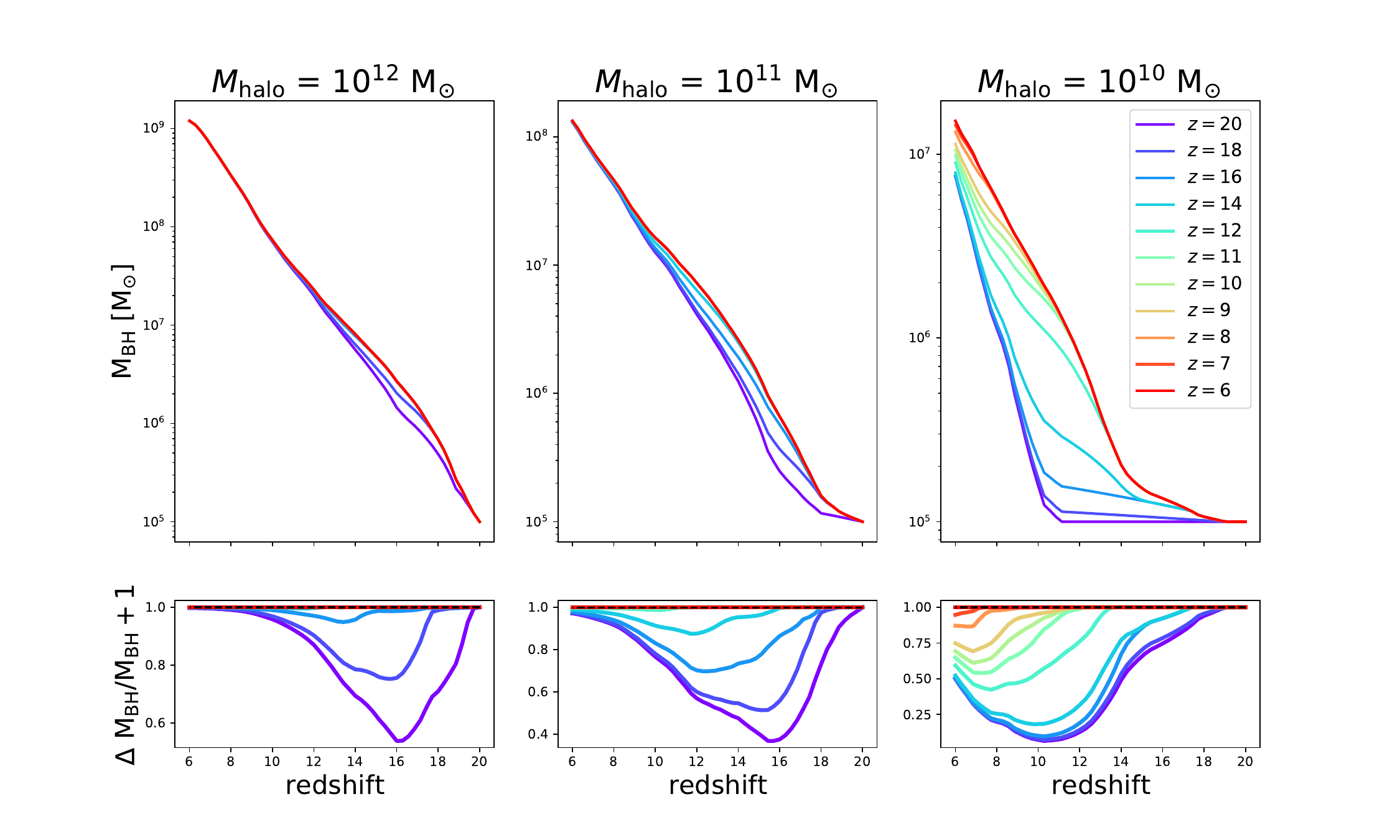}}\\
\end{center}
\caption{Just as Figure~\ref{fig:result3} but for our non-Eddington-limited model.}
\label{fig:result4}
\end{figure*}

\subsection{Extended reionization model}

While many of the parameters of reionization are still insufficiently constrained, there are a wealth of reionization models that estimate its timing and duration. Here we consider a set of instantaneous reionization models that we use to build an extended reionization model that estimates the volume-averaged effect of reionization on SMBH growth.

First, we consider a series of instantaneous reionization models at various redshifts. Because massive halos at early times are thought to be assembled in relatively dense regions of the cosmic web that are, on average, reionized early \citep{GM2022}, we consider reionization redshifts as high as $z=20$. Additionally, the assumption of instantaneous reionization for a local region is an apt approximation when considering the growth history of a single SMBH.

However, to evaluate the overall volume-averaged effect of reionization on SMBHs, we must consider the probability that a halo of a given mass will reside in a reionized region at a given time. Thus, we use 21cmFAST \citep{murray20,mesinger10} to generate a model of inhomogeneous reionization. 21cmFAST is a semi-numeric cosmological modeling tool that produces fast simulations of the early Universe using excursion set formalism with perturbation theory.

Our 21cmFAST calculations allow us to identify how halos are spatially distributed in reionized and non- reionized regions. Because in this model reionization occurs in the densest regions first, the largest halos are on average, reionized earliest. We run 21cmFAST with a standard cosmology and produce a halo field. We bin the halo masses and determine the ionization state of the location where each halo resides. Thus, we produce a function of redshift and halo mass that gives the probability that a halo resides in a reionized region. With this, we can determine what fraction halos of a given mass range are reionized within a given redshift interval. We can then use these fractions to produce a weighted mean of the different instantaneous reionization redshift models, ending up with a linear combination these reionization redshifts.

\subsection{Black hole growth model}
It has been long speculated that galaxy mergers can spur accretion onto a SMBH due to the funneling of gas to the center of the host galaxy. This has been studied extensively, with many observations suggesting that AGN are often associated with active mergers \citep{trakhtenbrot17,Benny2017,mathee23,Araujo2023,Perna2023}, though not all observational studies find similar trends \citep{Cisternas2011}. Theoretical models have also suggested that more frequent major mergers at high redshift can aid SMBHs in their rapid growth at early times \citep{Tanaka2014}. \citet{Hopkins2006} envisioned a "cosmic cycle" for galaxy formation in which mergers trigger AGN activity through gas inflows that result in SMBH growth. However, simulations have not always found mergers to be the sole or even dominant trigger of black hole growth \citep{Steinborn2018,Martin2018,Byrne2023,Ni22,Davies2023}.  Here we assume that major mergers do trigger BH growth and we explore the sensitivity of this BH growth to the minimum ratio of halo gas masses constituting a major merger. It should also be noted that as this minimum ratio decreases, this model corresponds more closely to one in which smooth accretion of all the gas falling into a halo continuously fuels black hole growth. 

To estimate how a black hole grows in response to a merger, we consider two simple models. 
In the first, Eddington-limited accretion is triggered by gas-rich mergers. Here we set the rate of growth to be the Eddington accretion rate for a duration equal to the gas dynamical timescale. The Eddington accretion rate is defined as
\begin{equation}
\dot{M}_{\rm BH} = \frac{4\pi G M_{\rm BH} m_{p}}{\epsilon c \sigma_{T}},
\end{equation}
where $M_{\rm BH}$ is the mass of SMBH as a function of time; $m_{p}$ is the proton mass; $\epsilon$ is the radiative efficiency, which we set to 0.1; and $\sigma_{T}$ is the Thomson scattering cross section. Note that this means the accretion rate is set by the mass of the SMBH. 

Our second model is inspired by previous semi-analytic models of galaxy formation that define the black hole accretion rate as a percentage of gas in a halo that is accreted over a given timescale. We use a prescription in which the black hole grows at the rate

\begin{equation}
\dot{M}_{\rm BH} = 0.01 M_{\rm halo} \frac{\Omega_b}{\Omega_{\rm DM}} / \tau
\end{equation}
where $\tau$ is either the dynamical timescale (for a dynamical timescale following a major merger), or the freefall timescale following the dynamical timescale after a merger. It is important to note that this means that this model is still merger driven, as the accretion rate following a merger far exceeds the baseline accretion rate. 

This prescription for black hole growth assumes that the black hole accretes one percent of the total gas mass in the progenitor halo over the given timescale. This is similar to the prescription in \citet{Bower2006}, who choose 0.5 percent of the stellar mass of the host halo's gas to be accreted onto the central black hole. Their choice of 0.5 percent of gas mass was selected to recover the normalization of the empirical local relation between galaxy bulge mass and black hole mass, whereas we have chosen a slightly higher value that is broadly in line with the estimated masses of SMBHs in the reionization era that lie above the local relation \citep{Pacucci2023}. This model also allows for super-Eddington growth, which may be more important for lower mass SMBHs at high redshift \citep{Shirakata2020}.

 We effectively assume a larger fraction of gas going into black holes than \citet{Bower2006}, in order to calibrate to the more rapid build up of SMBHs relative to stars in the highest-redshift galaxies, as compared to the lower-redshift galaxies to which \citet{Bower2006} calibrated their model. Assuming a star formation efficiency of 0.1, this corresponds to a factor of 20 times higher fraction of the gas accreted into the halo going into the black hole than assumed in \citet{Bower2006}.  This is broadly consistent with the black hole - stellar mass relation so far inferred from JWST observations in the epoch of reionization (e.g. \citealt{Pacucci2024, Li2024}; ).

In both of these black hole growth models, we set the seed black hole mass uniformly to $10^5$ M$_{\odot}$. This mass was selected so that the final SMBH masses were largely consistent with measured SMBH masses at high redshift. However, this choice is somewhat arbitrary as the seed mass will not change the fractional difference between different models. A $10^5$ M$_{\odot}$ seed is also consistent with predictions for direct collapse black hole seeds \citep{inayoshi19}, and it coincides with the lower mass end of JWST candidate black holes \citep{MaiolinoScholz2023}.  These $10^5$ M$_{\odot}$ black holes are seeded in our main progenitor halos at $z=20$. However, if we use a lower seed mass, our Eddington-limited model is unable to grow SMBHs rapidly enough to match the highest mass SMBHs found at high-z.

\subsection{Full model}

The full model is operated as follows:
\begin{enumerate}
    \item A merger tree calculation is run using \textsc{Galacticus}. The main progenitor's mass evolution is determined and all halos that merge directly into the main progenitor are identified.
    \item A black hole is seeded at $z=20$ in the main progenitor halo. We uniformly set the black hole seed mass to $10^5$ M$_{\odot}$.
    \item A gas fraction is assigned to each halo based on the redshift in which it merges into the main progenitor and the redshift of (local) reionization. Then a gas mass is calculated for each halo merging directly into the main progenitor halo. 
    \item Major mergers are identified based on the ratio of merging halos' gas mass to the gas mass of the main progenitor halo. We consider accretion episodes to be triggered by mergers with halo gas mass ratios of 0.3, 0.1 and 0.05, in order to test the sensitivity of our results to our assumption that major mergers are the triggers of BH growth.
    \item The dynamical timescale is calculated for the main progenitor halo at each redshift for which there is a major merger. Black hole accretion is initiated at the time of each major merger for the duration of the dynamical timescale. 
    \item We repeat the above process for a series of merger tree realizations with differing random seeds and then average the results. 
\end{enumerate}

\section{Results}
\label{sec:results}

\subsection{Early and late reionization models}
Our first model calculations involve an instantaneous "local" reionization. We compare an early reionization model with a late reionization model, where the early reionization occurs at $z=20$ and the late reionization occurs at $z=6$. Because we are considering the effect of reionization on an SMBH growing in a single dark matter halo, the instantaneous local reionization is a sufficient approximation. We compare the effect of an early and a late reionization in halos which grow to $10^{12}$, $10^{11}$, and $10^{10}$ M$_{\odot}$ by $z$ = 5.5. We also vary the threshold gas mass ratio that constitutes a "major merger". For these calculations, we assume major merger gas mass threshold ratios to be 1/10 and 1/20. Though the latter is quite low\footnote{Other studies define this ratio to be as high as 1/3 \citep{pearson19}.}, it is useful as an approximation of the inclusion of smooth accretion in our model. The reionization redshifts in these calculations are meant to bracket the most extreme cases of early and late reionization, in order to demonstrate an upper limit of what is possible within our model.



Figure~\ref{fig:result1} shows the growth histories of SMBHs in our Eddington-limited model. The lighter curves show the growth preceding a late reionization and the darker curves show growth following an early reionization. The dotted curves show an early reionization model with our weakest gas suppression condition, a threshold mass with a virial temperature of 1.0$\times 10^4$ K. The solid darker curves show our fiducial model, a threshold mass with a virial temperature of 2.5$\times 10^4$ K. The lower dashed curves show a calculation with our threshold halo mass that can retain gas following reionization as defined by one fourth of the Jeans mass. The upper panels show a final halo mass of $10^{12}$ M$_{\odot}$ at $z=5.5$, the middle panels show a final halo mass of $10^{11}$ M$_{\odot}$ at $z=5.5$ and the lower panels show a final halo mass of $10^{10}$ M$_{\odot}$ at $z=5.5$. The left panels show a condition where the gas mass of a "major merger" is defined as ten percent of the main progenitor's gas mass whereas the right panels show a relaxed threshold of five percent. The dashed black curve in the upper left panel shows constant, uninterrupted Eddington-limited growth. 

Figure~\ref{fig:result1} also demonstrates the limited capacity of merger-driven accretion episodes to grow SMBHs efficiently if we impose a stricter merger mass ratio and are limited to the Eddington rate.

The difference in the growth histories and final SMBH masses between early and late reionization are largely dependent on the condition used for the threshold mass below which halos cannot accrete following reionization. The difference is also largest for smaller progenitor halos where the merging halos are more likely to be below the threshold mass and at early times the progenitor halo itself may be below this threshold. While the fractional differences between the final masses do not vary significantly between the right and left panels, the final masses of all SMBHs are larger in the right-hand panels with a less strict condition for what consitutes a major merger. This is because there are more major mergers that trigger BH growth, despite the fact that reionization limits the gas mass in many of the halos which would otherwise undergo major mergers.

Figure~\ref{fig:result2} is the same as Figure~\ref{fig:result1} but for the growth histories of SMBHs in our non-Eddington-limited model. 
This model shows SMBH masses that tend to converge toward a similar final mass in our 10$^{12}$ M$_{\odot}$ and 10$^{11}$ M$_{\odot}$ halos. This is because the accretion rate in this calculation is proportional to the gas mass in the halo, which is relatively insensitive to the impact of reionization at late times when the main progenitor halo grows via mergers of relatively massive halos that are above our threshold mass for gas accretion during reionization.

\subsection{Effect of reionization redshift on SMBH growth}

While the preceding calculations show two extreme reionization redshifts, the following figures compare calculations of our SMBH growth histories with a range of reionization redshifts bracketed by $z=20$ and $z=6$.

Figure~\ref{fig:result3} shows the growth histories of SMBHs in our Eddington-limited model experiencing different local reionization redshifts and with gas accretion thereafter suppressed in halos with virial temperature below 2.5$\times$10$^{4}$ K. The lower panels show the fractional change in SMBH mass between the models assuming late ($z$ = 6) and earlier reionization in the upper panels. From left to right, the panels show our results for final halo masses of $10^{12}$ M$_{\odot}$, 10$^{11}$ M$_{\odot}$ and 10$^{10}$ M$_{\odot}$ at $z=5.5$. In this particular calculation, the gas mass of a "major merger" is defined as ten percent of the main progenitor's gas mass.

For the largest final halo mass we consider, 10$^{12}$ M$_{\odot}$, only the calculations with the earliest of reionization redshifts show significant SMBH growth suppression. A reionization redshift below $z = 16$ yields a final SMBH mass of at most ten percent lower than our calculation with a reionization redshift of $z = 6$. Meanwhile, for the case with a 10$^{11}$ M$_{\odot}$ halo we find a $z=11$ reionization redshift produces a final SMBH mass roughly twenty percent lower than our calculation with a reionization redshift of $z = 6$. Finally, our 10$^{10}$ M$_{\odot}$ halo experiences non-negligible SMBH growth suppression at redshifts as low at $z=7$. A reionization at $z=9$ yields a final SMBH mass that is $\sim$50 percent lower than a SMBH mass growing in a similar halo that experiences reionization at $z=6$. We find that the earliest reionization we consider ($z_{\rm reion} = 20$) may reduce a final super massive black hole's mass by [50, 70, 90] \% in halos of mass [$10^{12}$, $10^{11}$, $10^{10}$] M$_{\odot}$ by $z$ = 6.

Figure~\ref{fig:result4} shows the same as Figure~\ref{fig:result3} but for the growth histories of SMBHs in our non-Eddington-limited model. Unlike our Eddington-limited model, the final SMBH masses in the two more massive halos are mostly insensitive to the redshift of reionization. This is because the gas mass in these halos, taken to be proportional to the accretion rate in this model, is not significantly reduced due to reionization. This is because at later times the main progenitor halo is sufficiently massive. However, SMBH growth suppression is evident in the smaller progenitors of these halos at earlier times. 

\subsection{Extended reionization and the average effect over SMBH populations}

To address the cumulative impact of an extended and inhomogeneous reionization on the SMBH population, we calculate growth histories of SMBHs based on a set of weighted reionization redshifts. These weights are derived from a 21cmFAST calculation in which we find the percentage of halos at a given mass in a reionized region as a function of redshift. 

\begin{figure*}
\begin{center}
\resizebox{20.0cm}{!}{\includegraphics{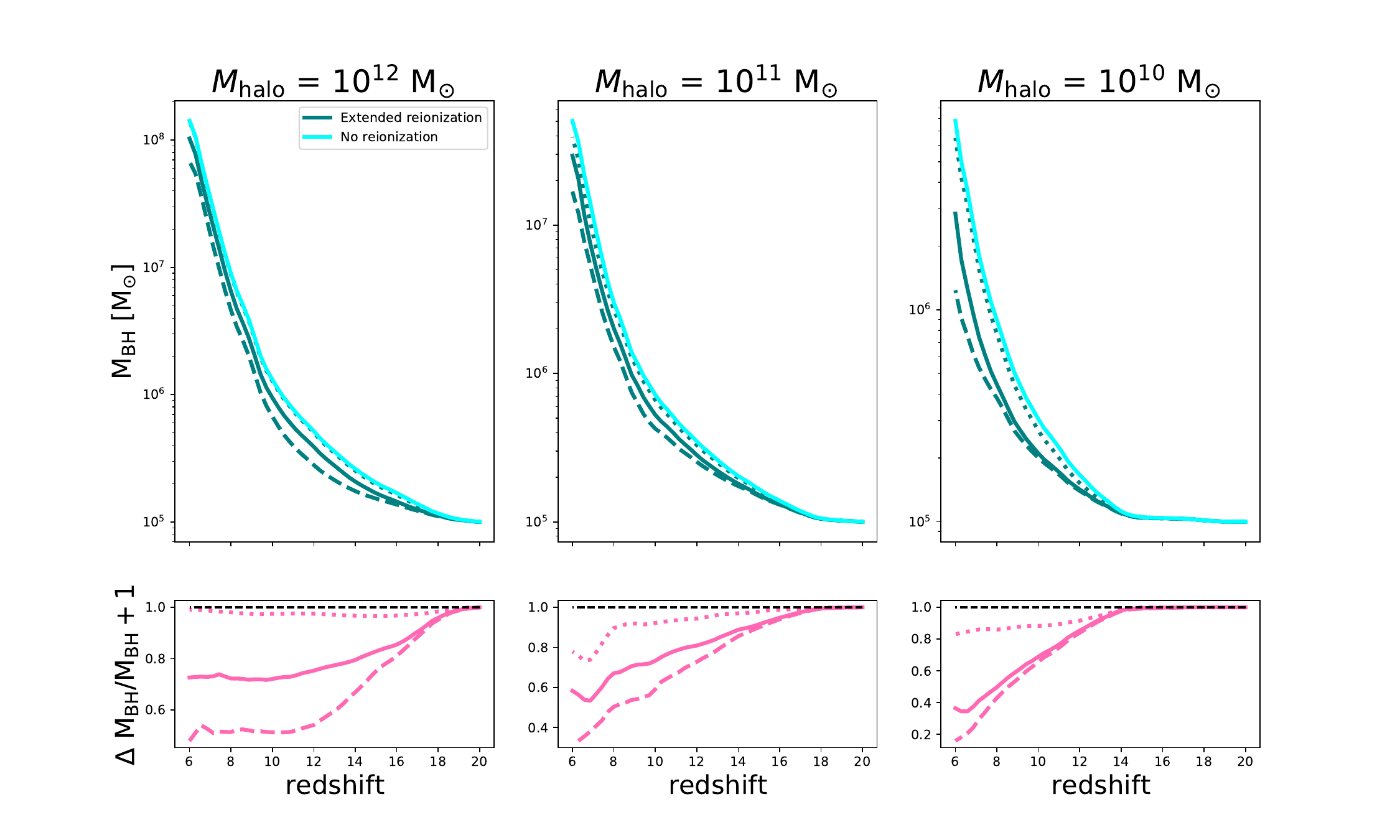}}\\
\end{center}
\caption{The growth histories of SMBHs in our Eddington-limited model with extended reionization. The darker curves show the average growth history of a SMBH affected by an extended reionization (as calculated by 21CMFast) and the light curves show the average growth history of a SMBH not affected by any reionization. The lower panels show the fractional change in SMBH mass between the two models in the upper panels. The left, middle and right panels show results for final halo masses of $10^{12}$ M$_{\odot}$, $10^{11}$ M$_{\odot}$, and $10^{10}$ M$_{\odot}$ respectively, at $z=5.5$. The dotted curves show the $10^4$ K virial temperature-derived threshold mass model, the darker solid curves show the $2.5\times10^4$ K virial temperature-derived threshold mass model, and the dashed curves show the $1/4$M$_{J}$-derived threshold mass model. In these calculations, the gas mass of a "major merger" that triggers BH growth is defined as ten percent of the main progenitor's gas mass.  
\label{fig:result5}}
\end{figure*}

Figure~\ref{fig:result5} shows the growth histories of SMBHs in our Eddington-limited model, with the darker curves showing the average growth history of a SMBH affected by an extended reionization (as calculated by 21cmFAST) and the light curves showing the average growth history of a SMBH not affected by any reionization. The lower panels show the fractional change in SMBH mass between the models assuming late ($z$ = 6) and earlier reionization in the two models in the upper panels. The left panels show results for the case of a final halo mass of $10^{12}$ M$_{\odot}$, the middle panels show those for a $10^{11}$ M$_{\odot}$ halo, and the right panels show a $10^{10}$ M$_{\odot}$ halo at $z=5.5$. As with Figures~\ref{fig:result1} and~\ref{fig:result2}, the dotted curves show the $10^4$ K virial temperature-derived threshold mass model, the darker solid curves show the $2.5\times10^4$ K virial temperature-derived threshold mass model, and the dashed curves show the $1/4$M$_{J}$-derived threshold mass model. In these models, the gas mass of a "major merger" that triggers BH growth is defined as ten percent of the main progenitor's gas mass.

While the average effect on SMBHs in 10$^{12}$ M$_{\odot}$ halos is roughly a modest twenty percent in our fiducial model, we find that SMBHs in halos of 10$^{11}$ M$_{\odot}$ may have average masses reduced by about fourty percent and reduced by more than half what one would expect without reionization in our 10$^{10}$ M$_{\odot}$ halo.\footnote{While we have presented the average BH masses produced in halos that grow to a given mass by $z$ = 5.5, we find that there is a significant spread of up to an order of magnitude in the BH masses due solely to variations in the merger histories of their host halos.} Because the regions hosting more massive halos are ionized earlier on average in our 21cmFAST calculation, the 10$^{12}$ M$_{\odot}$ host halo sees the local reionization redshifts weighted toward earlier times. Meanwhile, the smaller halos, though they experience more SMBH growth suppression for a given reionization redshift, have average local reionization redshifts that skew toward lower redshift.

\section{Discussion}
\label{sec:obs}

In general, we find a lower duty cycle in reionized regions, with SMBHs in regions reionized earliest experiencing the least overall accretion in comparison to their late-reionized counterparts. Intuitively, the effect of reionization is also most pronounced in smaller host halos. While these results do imply that reionization may affect early SMBH growth, it is important to understand how the limitations of our model may affect our results.

\subsection{Model limitations}
Our model is built on a key assumption, that mergers are a dominant driver of SMBH growth. While there are both theoretical findings \citep{Hopkins2006, Tanaka2014} and empirical evidence \citep{trakhtenbrot17,Benny2017,mathee23,Araujo2023,Perna2023} that mergers likely contribute to SMBH growth and AGN activity, there is limited consensus that mergers are the dominant contributor to SMBH growth \citep{Cisternas2011,Steinborn2018,Martin2018,Byrne2023,Davies2023}. Additionally, we do not explicitly consider how smooth accretion onto the progenitor halo may be affected by reionization, as ionizing radiation does affect the rate at which gas can cool and accrete from the IGM \citep{efstathiou92,dijkstra04,hambrick09,hambrick11}. Despite our not directly modeling smooth accretion, we do change the minimum ratio of what constitutes a major merger. As we make this condition less stringent, the lower mass major mergers may approximate a process similar to smooth accretion. 

We also do not consider black hole mergers. Presumably, many merging halos that host galaxies will host a central black hole as well. The growth of these black holes would, in principle, also be affected by reionization, especially as these halos by definition would be lower mass than the progenitor halo. 

It can also be seen in the previous section that our results are very sensitive to the temperature after reionization and how we calculate our threshold mass below which halos cannot accrete following reionization. Our simple models for gas suppression could clearly be improved upon by full cosmological simulations of the effect of reionization on low mass halos at very early times. Rather than a single threshold mass as a function of redshift, a model that considers partial gas occupation fractions could additionally increase our calculated impact of reionization. 

Finally, we have not explored whether the host galaxy's own radiation and SMBH feedback may be more important to the local intergalactic environment than the extragalactic background. In our model, we assume that all halos that will merge into the progenitor are reionized simultaneously and instantaneously, independent of the growth rate of the SMBH in our progenitor halo.  At earlier times or when the SMBH has had limited time to grow, this feedback may not yet have had sufficient time to ionize the local intergalactic environment. In fact, this is when we find that reionization will have the largest effect on the SMBH growth. Though radiation produced by a bright quasar itself will ionize the surrounding medium, the neutral gas in the quasar environment can also be retained if the ionizing radiation produced by the AGN is obscured, trapped \citep{begelman78,Inayoshi2016,johnson22,Ghodla2023}, or if the quasar's ionizing radiation does not yet extend far into the circumgalactic gas. Such obscured SMBH growth may explain the often small sizes of observed quasar proximity zones in epoch of reionization \citep{Satyavolu2023}.  In fact, recent JWST observations find a UV bright duty cycle of $\la$ 1 percent for quasars at $z$ $\sim$ 6, which suggests that most black hole growth during reionization may have been obscured \citep{Pizzati2024}.  This is also broadly consistent with independent data suggesting that radio-loud quasars at the highest redshifts are heavily obscured \citep{Capetti2024} and the discovery of a blazar at $z$ = 7 which suggests that most SMBH growth at high-z may be obscured and at super-Eddington rates \citep{Banados2024}. Additionally, these models can still be informative in the case where the AGN does reionize its local intergalactic environment, as they can provide a comparison between the growth of the SMBH prior to and after the feedback ionizing the local environment.

\subsection{Potential observables and measurements}

Due to JWST's accelerating pace of high-redshift SMBH discoveries \citep[e.g.][]{Uebler2023,Larson2023,Harikane2023,Matthee2023}, predictions of the effect of reionization on SMBH growth are more timely than ever. While measurements of the effect of reionization on SMBH growth are likely to be imprecise, in part due to the interplay of different feedback processes, it may be possible to uncover a general trend between SMBH masses and the ionization state of their intergalactic environments with a sufficiently large statistical sample. For instance, future measurements of both SMBH mass and the size of the ionized regions surrounding them, potentially including proximity zones produced by the AGN themselves, could provide an opportunity to uncover a signature of the effect of reioization on black hole growth.  Specifically, an anti-correlation between SMBH mass and the ionized region size would be consistent with reionization hindering SMBH growth.
A straightforward way to test this would be to (1) gather a statistical sample of high-redshift quasars and (2) obtain high-resolution spectra of individual quasars at high redshift to measure their proximity zones or damping wings.

Ground based optical--near-infrared (NIR) spectrographs have measured quasar proximity zones at $z\sim5.8$--6.5, which can constrain both the neutral fraction of the IGM as well as the quasar lifetimes \citep[e.g.][]{fan06,Willott2007,Carilli2010,Eilers2017,Satyavolu2023}.
Another probe of the neutral fraction, the quasar damping wing can be observed from the ground at $z\sim 7-7.5$ \citep[e.g.][]{Mortlock2011,Greig2017,Greig2019,Davies2018,Wang2020}, so detailed observations of the local quasar environment are already in reach for quasars during that stage of reionization \citep{yang21,Fan2023}.
The Near InfraRed Spectrograph (NIRSpec) aboard the JWST \citep{Jakobsen2022} can now carry out quasar spectroscopy from 1.6--5$\mu$m. This is ideal for probing both the proximity zones or damping wings of the highest redshift quasars, as well as accurate measurements of SMBH masses and accretion rates from the broad Balmer lines \citep{Eilers2023,Marshall2023,Yang2023}. Having both IGM measurements and accurate BH properties will allow for a more complete understanding of the interplay between reionization and SMBH growth, and allow us to investigate whether, for example, there is a trend of lower SMBH masses in halos having reionized earlier. These efforts will be complicated by the presence of observational biases and measurement uncertainties, and so a  significantly larger sample of high-z quasars would be required, particularly for quasars at $z>7$ where less than 10 are currently known.

A significant number of SMBHs at $7<z<9$ are expected to be discovered with the recently launched EUCLID mission \citep{Euclid2019}, an optical--NIR space telescope. By mapping a large area of the sky ($>15,000 \rm{deg}^2$ to 24 mag), Euclid is expected to discover more than 100 luminous quasars at $7 < z < 8$ \citep{Laureijs2011,Barnett2019}, ideal for studying the most massive early SMBHs.
Additionally, the NIR {\it Roman Space Telescope} \citep{Spergel2015} will dramatically increase the number of high-redshift quasars discovered, isolating much smaller, less luminous black holes by probing depths of $\sim27$mag across $\sim2,000 \rm{deg}^2$ with the High Latitude Wide Area Survey.
The {\it Rubin Observatory} \citep{Ivezic2019} Legacy Survey of Space and Time (LSST)  will provide deep optical imaging which will be a key complement to these space telescopes, particularly for dropout identification and source classification.
Roman and Rubin combined are predicted to increase the current known number of high-z quasars by at least a factor of 3, and to detect quasars out to z>8.5 \citep{Tee2023}.
Thus over the next decade, we expect to have a large statistical sample of high-z quasars.
JWST and other upcoming missions, specifically the ELTs, will have the capabilities to measure SMBH masses and accretion rates of these new quasars with deep spectroscopy \citep{fan19}.

It may be that the sweet spot to test the effect of reionization on BH growth is at the low-mass end of observed high-z BH mass function, because (1) we've shown that it is the smallest host halos in which the effect of reionization is greatest in limiting BH growth and (2) lower-mass BHs are less likely to produce ionizing radiation that dominates the local intergalactic radiation field, yielding measurements of their proximity zones more likely to be indicative of the full local reionization history.

Eventually, 21 cm experiments will be an important probe of the reionization epoch quasar environment. Not only will these experiments have the ability to track the global \HI\ fraction and reionization directly, they will also be able to detect and measure the properties of the gaseous environment around growing SMBHs that exist during reionization. While these observations may still be a decade away, they will answer many of the key questions that remain about reionization and its impact on galaxy formation and SMBH growth.

Finally, the Lynx X-ray Observatory is a proposed mission that claims to have the ability to probe accreting seed-mass BHs at $z=10$ \citep{lynx18}.  Because we find that the BH mass decrement due to reionization may be the largest at such high redshifts, particularly in the lowest-mass host halos (see Figure~\ref{fig:result4}), such observations would be particularly valuable for testing our predictions.

\section{Summary}
\label{sec:summ}
We have estimated the impact of reionization on the growth of SMBHs using a semi-analytic model. Our calculation combines merger trees with a series of reionization models to estimate the reduction in gas-rich mergers that trigger SMBH growth. Using these calculations in tandem with black hole growth models, we find that early reionization may reduce a SMBH's mass by [50, 70, 90] \% in halos of mass [$10^{12}$, $10^{11}$, $10^{10}$] M$_{\odot}$ by $z$ = 6.

We investigate the impact of local reionization redshift on SMBH growth histories and find that for the most massive halos we consider (10$^{12}$ M$_{\odot}$), only the calculations with the earliest of reionization redshifts show perceptible SMBH growth suppression. In particular, we find that reionization below $z = 16$ yields a final SMBH mass of at most ten percent lower than our calculation with a reionization redshift of $z = 6$. On the other hand, an SMBH in a 10$^{11}$ M$_{\odot}$ halo with a $z=12$ reionization redshift produces a final SMBH mass roughly twenty percent lower than our calculation with a reionization redshift of $z = 6$. Most dramatically, an SMBH in a 10$^{10}$ M$_{\odot}$ halo experiences noticeable growth suppression at redshifts as low at $z=7$. We find that in a 10$^{10}$ M$_{\odot}$ halo, reionization at $z=9$ yields a final SMBH mass that is $\sim$50 percent lower than a SMBH mass growing in a similar halo that experiences reionization at $z=6$.

Finally, we estimate the total impact of an extended reionization on the global population of SMBHs in halos of 10$^{10}$ M$_{\odot}$, 10$^{11}$ M$_{\odot}$, and 10$^{12}$ M$_{\odot}$ at $z$ = 5.5. While the overall effect is a modest twenty percent average reduction in SMBH mass in 10$^{12}$ M$_{\odot}$ halos, forty percent average reduction in SMBH mass in 10$^{11}$ M$_{\odot}$ halos, we find that SMBHs in halos of $\la$ 10$^{10}$ M$_{\odot}$ may have average final masses reduced by greater than a factor of two below what one would expect without reionization. 

We note that our model does not incorporate galactic feedback processes and does not explicitly model AGN feedback. However, there is growing evidence that the neutral gas in the greater environment can be retained in cases where radiation produced by the AGN is obscured or trapped, as JWST observations are suggesting may be typical. In cases where the AGN feedback is responsible for reionizing the local intergalactic environment, the SMBH undergoes an initial growth phase before this reionization, potentially leading to a more rapid build-up during its early stages.

Ultimately, we find a lower duty cycle in earlier reionized regions, however the extent of this effect is very dependent on model parameters.  In particular, we find that the extent of the suppression of SMBH growth due to reionization is critically dependent on the extent to which gas is suppressed in low mass halos at different redshifts, and thus on the temperature following reionization, as well as on whether or not black hole growth is limited to the Eddington rate. That said, we also find that growing SMBHs through major mergers with Eddington-limited accretion does not appear to be able to match the most massive empirical SMBH masses at high redshift.

These predicted effects may be tested with the census of SMBHs that will be taken by current and future facilities that will add to the ground-breaking high-z data on early SMBH growth that has been provided by the JWST.

\begin{acknowledgments}
P. R. U. S. would like to acknowledge support of an LDRD Director’s Postdoctoral Fellowship at Los Alamos National Laboratory (20210942PRD2).  The authors are also grateful to Andrew Benson for invaluable assistance with \textsc{Galacticus}, and to Anson D'Aloisio and Christina Eilers for formative discussions.

\end{acknowledgments}

\bibliography{References}


\end{document}